\documentclass{elsart}
\usepackage[T1]{fontenc}
\usepackage[latin1]{inputenc}
\usepackage{graphicx}

\usepackage{amssymb}

\begin{document}

\begin{frontmatter}

	\title{Universal Scaling in Saddle-Node Bifurcation Cascades (II) \\ Intermittency Cascade }

	\author[addr1,addr2]{Jesús San-Martín}
	\address[addr1]{Departamento de Matemática Aplicada, E.U.I.T.I, Universidad Politécnica de Madrid. Ronda de Valencia 3, 28012 Madrid Spain}
	\address[addr2]{Departamento de Física Matemática y Fluidos, U.N.E.D. Senda
del Rey 9, 28040 Madrid Spain}
	\ead{jsm@dfmf.uned.es}

	\begin{abstract}
The presence of saddle-node bifurcation cascade in the logistic equation
is associated with an intermittency cascade; in a similar way as a
saddle-node bifurcation is associated with an intermittency. We merge
the concepts of bifurcation cascade and intermittency. The mathematical
tools necessary for this process will describe the structure of the
Myrberg-Feigenbaum point.
	\end{abstract}
	

\end{frontmatter}

\section{Introduction}

Period-doubling \cite{Feigenbaum78,Feigenbaum79}, Quasi-periodicity
\cite{Ruelle71,Newhouse78} and Intermittency \cite{Manneville79,Pomeau80}
are well known routes of transition from periodic to chaotic behavior,
and whose origin is in local bifurcations. Initially, the system has
a stable limit cycle, for a range control parameter $r$. As this
parameter is increased beyond a critical value $r_{c}$ system behavior
changes according to a local bifurcation that occurs at $r_{c}$.

In order to study the genesis of the transition we resort to the Poincare
section. In this section, the original stable limit cycle of the system
generates a fixed point, whose evolution is studied in parallel to
the control parameter changes.

Paying attention to the different local bifurcation kinds we will
have different transitions to chaos. So, if the fixed point shows
successive pitchfork bifurcations, which double repeatedly the period
of the original orbit, the Feigenbaum period doubling cascade is obtained.
The final ending is a period-$\infty$ orbit, a chaotic attractor.

Quasi-periodicity occurs as a new Hopf bifurcation generates a second
frequency in the system, which is incommensurate with the original
system frequency. If an irrational winding number is fixed it goes
through chaos.

At last, the intermittency is a chaotic regime characterized by an
apparently regular behavior, which undergoes irregular bursts from
time to time. This intermittency between two behaviors names this
kind of chaos.

The regular behavior or laminar regime corresponds to an evolution
of the system in a narrow region or a channel in the phase space.
Such regular behavior stems from the fact that the system maintains
a {}``ghost'' of laminar regime

Whereas in the other transitions to chaos, Period-doubling and Quasi-periodicity,
the system is totally regular before the transition and chaotic later.
This does not happen with the intermittency. The intermittency shows
a continue transition from regular behavior to a chaotic one. The
smaller the value $\varepsilon=r-r_{c}$ is the longer the laminar
regime will be and the lesser it will be altered by a chaotic behavior,
due to the fact that the average time of laminar regime $\left\langle l\right\rangle $
for small $\varepsilon$ is $\varepsilon^{-\beta}$, being $\beta>0$.
Therefore, beyond $r_{c}$the laminar behavior alternates with irregular
bursts, the smaller (bigger) the value $\varepsilon$ is the more
(less) the laminar regime $\left\langle l\right\rangle $ is.

The intermittency transition was discussed by Pomeau and Manneville
\cite{Manneville79,Pomeau80}. They pointed out three kinds of intermittency.
The system is periodic for parameter values smaller than the critical
point $r<r_{c}$. This periodic behavior generates a stable fixed
point in the Poincare section. As the parameter reaches the critical
value $r=r_{c}$the fixed point losses its stability. The loss of
stability is caused when the eigenvalue modulus of the linearized
Poincare map becomes larger than unit. This may happen is in three
different ways.

i) There is a real eigenvalue crossing the unit circle by plus one.
A saddle-node bifurcation is generated, which has associated type-I
intermittency. In this case the average length of laminar regime time
is $\left\langle l\right\rangle \sim\varepsilon^{-\frac{1}{2}}$.

ii) A couple of complex conjugate eigenvalues crosses the unit circle.
This circumstance is associated with the birth of a Hopf bifurcation
and it involves a type-II intermittency with $\left\langle l\right\rangle \sim\varepsilon^{-1}$

iii) A real eigenvalue crosses the unit circle by minus one, generating
a flip bifurcation. This time the type-III intermittency occurs and
we obtain $\left\langle l\right\rangle \sim\varepsilon^{-1}$.

Other kinds of intermittencies have been studied such as type-X \cite{Anil97}
one which shows a transition with hysteresis and type-V one \cite{Bauer92}
for discontinuous maps. Intermittencies have also been studied whose
laminar regime occurs alternatively between two channels, as a result
of symmetry in the problem \cite{SanMartin99}. 

The generalization and logical development of this latter problem
is to find intermittencies with an arbitrary number of channels. Furthermore
it is desirable to have a different number of channels for different
values of control parameters in the same system, and not to look for
different systems \emph{ad hoc} with the appropriate symmetries which
show a fixed number of channels. We say that such behavior is desirable
in an unique system because, if a change of the control parameter
implies an increase of the number of channel, what is obtained is
an intermittency cascade; in a similar way a change of the control
parameter in the logistic map generates a period doubling cascade.

In this paper we are going to show that such phenomenon occurs in
the logistic map, benefiting from the fact that this map shows saddle-node
bifurcation cascades \cite{SanMartin05} and that the type-I intermittency
is associated with the saddle-node bifurcation. We will characterize
control parameter values at which successive intermittencies are generated,
how some intermittencies are related to others, the number of channels,
which is the average time of laminar regime of intermittencies, and
what relationship there is between such regimes for different intermittencies.
To answer these questions we will use the universal properties of
the logistic map \cite{Feigenbaum78,Feigenbaum79} and the saddle-node
bifurcation cascade this map shows \cite{SanMartin05}.

The saddle-node bifurcation cascade is a sequence of saddle-node bifurcations
in which the number of fixed points showing this kind of bifurcation
is duplicated. The successive elements of the sequence are given by
an equation identical to the one that Feigenbaum found for period
doubling cascade \cite{SanMartin05}.

The way of acting is as follows. As we mentioned above type-I intermittency
is associated to one saddle-node bifurcation. Therefore, each element
of the sequence of the saddle-node bifurcation cascade has associated
a type-I intermittency. The number of channels of this intermittency
coincides with the number of fixed points that simultaneously show
a saddle-node bifurcation. For instance, the saddle-node bifurcation
cascade, symbolized by the sequence $3$,$3\cdot2$,$3\cdot2^{2}$,$3\cdot2^{3}$,...,$3\cdot2^{q}$,
points out that there are $3$ fixed points at a first parameter value
$r=r_{3}$, there are $3\cdot2$ at $r=r_{3\cdot2}$ and so on. Each
saddle-node fixed point contributes to the intermittency with one
channel, accordingly in this intermittency cascade there will be a
sequence of intermittencies with $3$,$3\cdot2$,$3\cdot2^{2}$,$3\cdot2^{3}$,...,$3\cdot2^{q}$,....
channels.

The channels responsible for laminar regime are close to critical
points of logistic map (Fig. \ref{cap:fig2}). The way, how the neighborhood
of these critical points are contracted in the successive iterated
of the map, determines how the channels are contracted and it allows
us to look for the connection between them. The scaling of the neighborhood
of critical points for iterated map was calculated near a pitchfork
bifurcation by Feigenbaum \cite{Feigenbaum78}. We will follow this
work to calculate the scaling near a saddle-node bifurcation, because
it is here where intermittency occurs.

As many iterated one-dimensional maps are nearly quadratic under renormalization
\cite{Gukenheimer87}, we have to expect that intermittencies cascade
is a common phenomenon in many natural processes. 

Let be the logistic equation $x_{n+1}=f(x_{n})=r\, x_{n}(1-x_{n})$.
The graph of $f^{3}$ ($f^{n}=f\circ\quad\circ f$) is shown in Fig.
\ref{cap:fig1}, where it is tangent to line $x_{n+1}=x_{n}$, at
$r=r_{c}=1+\sqrt{8}$ , which means a period-$3$ orbit exists . There
are three saddle-node fixed points, and we are at the genesis of a
saddle-node bifurcation. For $r>r_{c}$ the valleys and the hills
of $f^{3}$ are sharper than at $r=r_{c}$, and each saddle-node point
has generated two points: one saddle and one node. If we decreased
from $r>r_{c}$ to $r<r_{c}$ we would observe that the saddle point
approaches the node one, touching at $r=r_{c}$, as the saddle-node
bifurcation occurs. For $r<r_{c}$ the valleys and the hills are pulled
away from the diagonal and saddle-node fixed points disappear. After
the bifurcations have disappeared three narrow channels, delimited
by the graph of $f^{3}$ and the diagonal, remain, which are responsible
for laminar regime of intermittency. This is what we meant when we
said earlier that each S-N point would be responsible for generating
a channel in the intermittency cascade.

Each iterated of logistic map will take a long time to go through
these channels (see Fig. \ref{cap:fig2}). The average number of iterated
inside a channel is given by $\left\langle l\right\rangle \sim\varepsilon^{-\frac{1}{2}}$\cite{Pomeau80},
and so it is for the set of three channels.

The saddle-node bifurcation cascade involves saddle-node bifurcations
with $q$,$q\cdot2$,$q\cdot2^{2}$,$q\cdot2^{3}$,...,$q\cdot2^{n}$,
$q\neq2^{m}$ fixed points for the maps $f^{q}$,$f^{2q}$,....,$f^{q2^{n}}$
respectively, and the same number of channels for the intermittency.
Fig. \ref{cap:fig3} and \ref{cap:fig4} show saddle-node bifurcations
for $f^{3\cdot2}=f^{6}$and $f^{3\cdot2^{2}}=f^{12}$ respectively.

We want to connect the length of laminar regime $\left\langle l\right\rangle $
of one intermittency with the laminar regime of other intermittencies
present in the cascade. If we notice in the neighborhood of point
$\left(\frac{1}{2},\frac{1}{2}\right)$in Fig. \ref{cap:fig3} we
will see that the graph of $f^{3}$ is reproduced in Fig. \ref{cap:fig1}
at $r=r_{c}$, escalated by a factor $\frac{1}{\alpha}$, $\alpha>1$.
The same can be said in the neighborhood of point $\left(\frac{1}{2},\frac{1}{2}\right)$in
Fig. \ref{cap:fig4}. As the iterated $f^{q2^{n}}$ with $n\longrightarrow\infty$
are considered the constant $\alpha$ we get is the Feigenbaum constant.
(see appendix)

If we come back to Fig. \ref{cap:fig3} we will notice that $f^{3\cdot2}$reproduces
again the graph of $f^{3}$ in the neighborhood of $\left(f(\frac{1}{2}),f(\frac{1}{2})\right)$,
and that this one is not escalated by the same factor as in the neighborhood
of the point $\left(\frac{1}{2},\frac{1}{2}\right)$. From one iterated
to the following one , that is, from $f^{3\cdot2^{n}}$to $f^{3\cdot2^{n+1}}$
, half of the neighborhood of critical points escalate approximately
with $\frac{1}{\alpha}$ and the other half with $\frac{1}{\alpha^{2}}$
(see appendix). We will be able to answer our questions because $f^{3}$
is reproduced in the neighborhood of critical points of $f^{3\cdot2^{n}}$and
also because we know how these neighborhoods scale in the successive
elements of the cascade

\section{Intermittency Cascade}

Let $r_{q\cdot2^{n},SN}$ be the parameter value at which $f^{q\cdot2^{n}}$,
$q\neq2^{m}$ has a saddle-node bifurcation, that is, $f^{q2^{n}}$has
a saddle-node orbit with $q\cdot2^{n}$points. The points of this
orbit are located right where the function $f^{q2^{n}}$ is tangent
to the line $y=x$. The $q\cdot2^{n}$points can be classified into
$2^{n}$ sets, each one of them having $q$ points. The $2^{n}$ sets
correspond to $2^{n}$ critical points of $f^{2^{n}}$closest to the
line $y=x$. The $q$ saddle-node points are captured in a neighborhood
of each one of these critical points, in other words, the graph of
$f^{q}$ is captured in every one of such neighborhoods; for instance,
in $f^{3\cdot2^{2}}$we notice how the graph of $f^{3}$ is captured
in the neighborhoods of the $2^{2}$ critical points of $f^{2^{2}}$
(Fig. \ref{cap:fig4}).

If we choose \begin{equation}
r=r_{q\cdot2^{n},SN}-\varepsilon\,\,\,\,0<\varepsilon\ll1\label{r}\end{equation}
then the saddle-node bifurcation will be about to occur. In these
conditions, there are $q\cdot2^{n}$ points where $f^{q2^{n}}$ is
almost tangent to the line $y=x$. In such points there are $q\cdot2^{n}$
channels, which are formed by the graph of $f^{q2^{n}}$ and the line
$y=x$. These channels are the narrower the smaller $\varepsilon$
is in Eq. (\ref{r}), and in them the laminar regime occurs. The time
to cross the channel depends on $\varepsilon$.

Let $\left\langle l\right\rangle _{n}$be the average time that the
iterates of $x_{n+1}=f(x_{n})$ spend to cross the $q\cdot2^{n}$
channels generated by $f^{q2^{n}}$. If we consider the laminar regime
of an intermittency of $f^{q\cdot2^{n+1}}$ then the number of channels
will become duplicated because $f^{q\cdot2^{n+1}}=f^{q\cdot2^{n}}\circ f^{q\cdot2^{n}}$,
in other words, the graph of $f^{q}$ is duplicated close to the critical
points of $f^{2^{n}}$. But in the doubling process the replicas of
graph of $f^{q}$ are contracted, half of them as $\frac{1}{\alpha}$
and the others as $\frac{1}{\alpha^{2}}$ as $n\rightarrow\infty$,
where $\alpha$ is the Feigenbaum constant (see appendix). Accordingly,
for $\varepsilon=r_{q\cdot2^{n+1},SN}-r$ the intermittency of $f^{q\cdot2^{n+1}}$
shows the channels of the intermittency of $f^{q\cdot2^{n}}$ duplicated,
but half of them contracted as $\frac{1}{\alpha}$ and the other contracted
as $\frac{1}{\alpha^{2}}$. As the average time for the intermittency
of $f^{q\cdot2^{n}}$is $\left\langle l\right\rangle _{n}$ it turns
out that the average time for the intermittency of $f^{q\cdot2^{n+1}}$
will be $\frac{\left\langle l\right\rangle _{n}}{\alpha}$, which
comes from the channels contracted by $\frac{1}{\alpha}$, plus $\frac{\left\langle l\right\rangle _{n}}{\alpha^{2}}$,
which are given by the channels contracted by $\frac{1}{\alpha^{2}}$.
In conclusion, the average time of laminar regime of the intermittency
of $f^{q\cdot2^{n+1}}$ is $\left\langle l\right\rangle _{n}(\frac{1}{\alpha}+\frac{1}{\alpha^{2}})$,
where $\left\langle l\right\rangle _{n}$is the average time of laminar
regime of $f^{q\cdot2^{n}}$, and both $f^{q\cdot2^{n}}$ and $f^{q\cdot2^{n+1}}$are
at the same distance $\varepsilon$from the corresponding saddle-node
bifurcation in the parameter space, that is, $r=r_{q\cdot2^{n+1},SN}-\varepsilon$
and $r=r_{q\cdot2^{n},SN}-\varepsilon$.

Let's notice that in a saddle-node bifurcation cascade the laminar
regime from an intermittency to the next one in the sequence is decreased
by a factor $(\frac{1}{\alpha}+\frac{1}{\alpha^{2}})$. Accordingly,
the average time for intermittency of $f^{q\cdot2^{n+m}}$ is\begin{equation}
\left\langle l\right\rangle _{n+m}=\left\langle l\right\rangle _{n}(\frac{1}{\alpha}+\frac{1}{\alpha^{2}})^{m}\label{long-n-m}\end{equation}

Given the saddle-node bifurcation cascade of $f^{q\cdot2^{n}}$,$f^{q\cdot2^{n+1}}$,....,$f^{q\cdot2^{n+m}}$,..
if the bifurcation parameter of $f^{q\cdot2^{n}}$is at $r_{q\cdot2^{n},SN}$
then the other bifurcation parameters are given by \cite{SanMartin05}

\begin{equation}
r_{q\cdot2^{n+1},SN}=\frac{1}{\delta}r_{q\cdot2^{n},SN}+(1-\frac{1}{\delta})r_{\infty}\label{Fei-Jesus}\end{equation}
where $\delta$is the Feigenbaum constant; and $r_{\infty}$is the
Myrberg-Feigenbaum point of a canonical window where Feigenbaum cascade
finishes and where also all saddle-node bifurcation cascades finish,
whatever $q\neq2^{m}$ is.

Eqs. (\ref{long-n-m}) and (\ref{Fei-Jesus}) determine the intermittency
cascade, because the parameter values at which it occurs and the average
time of its corresponding laminar regimes are known.

The former results are valid if a intermittency cascade occurs in
a period-$j$ window instead of a canonical window. Because both the
scaling of laminar regime $(\frac{1}{\alpha}+\frac{1}{\alpha^{2}})$
and Eq. (\ref{Fei-Jesus}) are valid in a period-$j$ window, although
the Eq. (\ref{Fei-Jesus}) turns into\[
r_{q\cdot2^{n+1},SN}=\frac{1}{\delta}r_{q\cdot2^{n},SN}+(1-\frac{1}{\delta})r_{\infty,j}\]
to indicate that the convergence is the one to the Myrberg-Feigenbaum
point $r_{\infty,j}$ of the period-$j$ window.

There is a second way of changing the control parameter in the intermittency
cascade, which is more important to the experimenters.

The Eq. (\ref{long-n-m}) gives the length of average times to an
intermittency cascade associated with the saddle-node bifurcation
cascade of $f^{q\cdot2^{n}}$,$f^{q\cdot2^{n+1}}$,....,$f^{q\cdot2^{n+m}}$,..,
if the value of $\varepsilon$ is constant. Such value gives the distance
from the control parameter to the saddle-node bifurcation parameter.
Nonetheless, it is possible to change $\varepsilon$in the successive
saddle-node bifurcations in such a way that the average time of laminar
regimen is kept constant, and equal to a $\left\langle l\right\rangle _{n}$,
for the whole intermittency cascade. For that purpose, all we need
is to have the value $\varepsilon$, taken in the first intermittency,
is rescaled by a factor $(\frac{1}{\alpha}+\frac{1}{\alpha^{2}})$
for each one of the successive intermittencies of the cascade, that
is, the values\begin{equation}
\varepsilon,\varepsilon(\frac{1}{\alpha}+\frac{1}{\alpha^{2}}),...,\varepsilon(\frac{1}{\alpha}+\frac{1}{\alpha^{2}})^{m},...\label{epsilones}\end{equation}
for $m=0,1,2,3,...$. This is so because \begin{equation}
\left\langle l\right\rangle \propto\frac{1}{\varepsilon}\label{eq:lm}\end{equation}
 for the type-I intermittency, which is present in the logistic equation.
If the values of (\ref{epsilones}) are introduced in Eq. (\ref{eq:lm})
then the laminar regimen is increased in a factor which is the same
as the one that contracts according to Eq. (\ref{long-n-m}). The
result is that the average time of laminar regimen stays constant.

It is necessary to change $\varepsilon$ in this way. As Eq. (\ref{Fei-Jesus})
shows a geometric progression, of ratio $\frac{1}{\delta}$, if we
held $\varepsilon$ constant then very quickly the value $r=r_{q\cdot2^{n},SN}-\varepsilon$
would not be within the parameter interval $\left[r_{q\cdot2^{n+m+1},SN},r_{q\cdot2^{n+m},SN}\right]$and
we would not observe channels corresponding to two successive saddle-node
bifurcation of the cascade. The result would be that $r\ll r_{q\cdot2^{n+m+1},SN}$
and the intermittency cascade would not be observed. Obviously, this
is vital for the experimenters and for the development of numerical
experiments as well.

Bear in mind that as $\varepsilon$changes as in Eq. (\ref{epsilones})
it turns out that also $\varepsilon$ changes as geometric progression
of ratio $(\frac{1}{\alpha}+\frac{1}{\alpha^{2}})$. This geometric
progression converges faster than the progression Eq. (\ref{Fei-Jesus}).
Accordingly, the parameter value $r$ can be held such that $r\in\left[r_{q\cdot2^{n+m+1},SN},r_{q\cdot2^{n+m},SN}\right]$,
and therefore the channel associated with the saddle-node bifurcation
at $r_{q\cdot2^{n+m},SN}$ can be observed. It is necessary for the
experimenter to change the parameter as in Eq. (\ref{epsilones}),
in order to stay close to successive saddle-node bifurcations of the
cascade and get a constant value of $\left\langle l\right\rangle $.
It is easy to get this variation because the bifurcation parameters
are given by Eq. (\ref{Fei-Jesus}).

\section{Myrberg-Feigenbaum point structure}

\begin{figure}[ppp]

\caption{\label{cap:fig1}Saddle-Node bifurcation genesis for $f^{3}$: $r<r_{c}$
pre-bifurcation (thin dashed line), $r=r_{c}$ at bifurcation (thick
dashed line, tangent to $y=x$), $r>r_{c}$ post-bifurcation (dotted
line).}

\begin{center}\includegraphics[%
  width=0.70\textwidth,
  keepaspectratio,
  angle=-90]{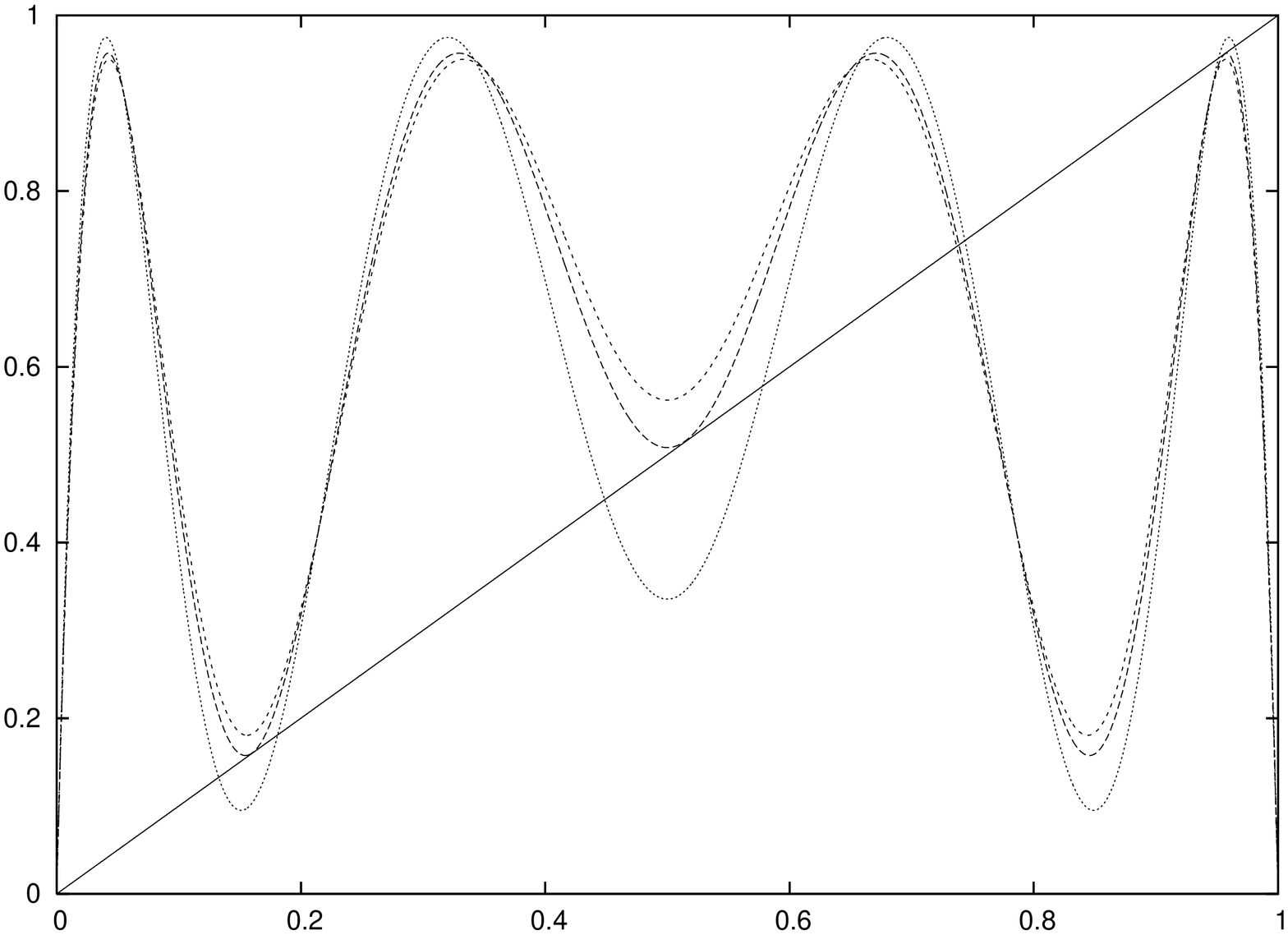}\end{center}
\end{figure}

\begin{figure}[ppp]

\caption{\label{cap:fig2}Thick line shows that many iterations are necessary
to cross the channels near a Saddle-Node bifurcation point.}

\begin{center}\includegraphics[%
  width=0.70\textwidth,
  keepaspectratio,
  angle=-90]{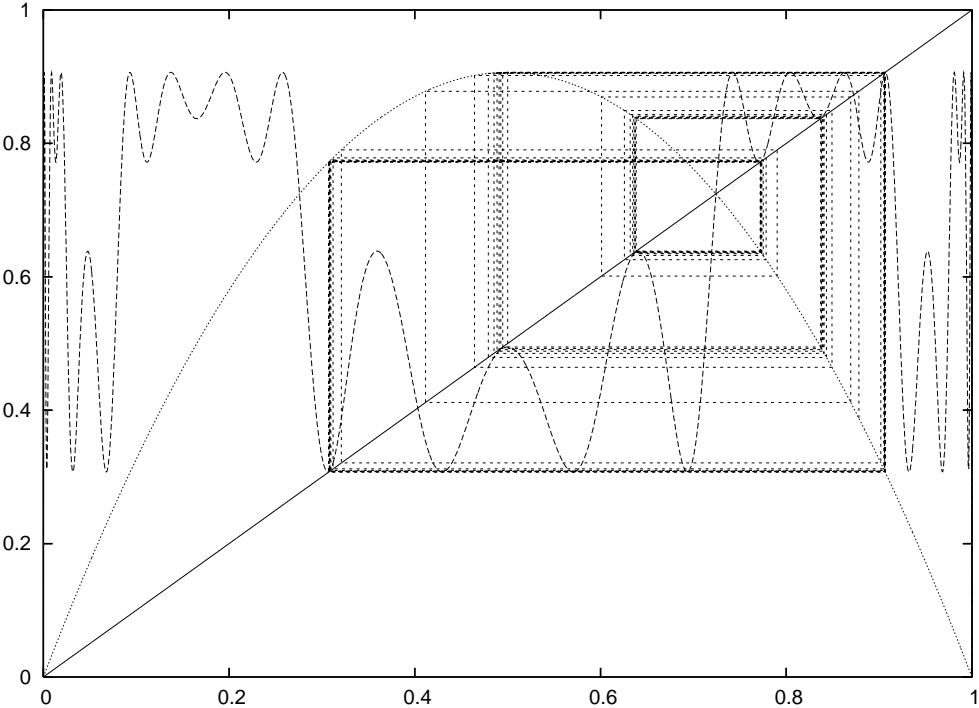}\end{center}
\end{figure}

\begin{figure}[ppp]

\caption{\label{cap:fig3}Graph of $f^{3\cdot2}$(thin dashed line), at Saddle-Node
bifurcation, reproduces the graph of $f^{3}$ (see Fig. \ref{cap:fig1})
around $f^{2}$ (thick dashed line) maxima and minimum.}

\begin{center}\includegraphics[%
  width=0.70\textwidth,
  keepaspectratio,
  angle=-90]{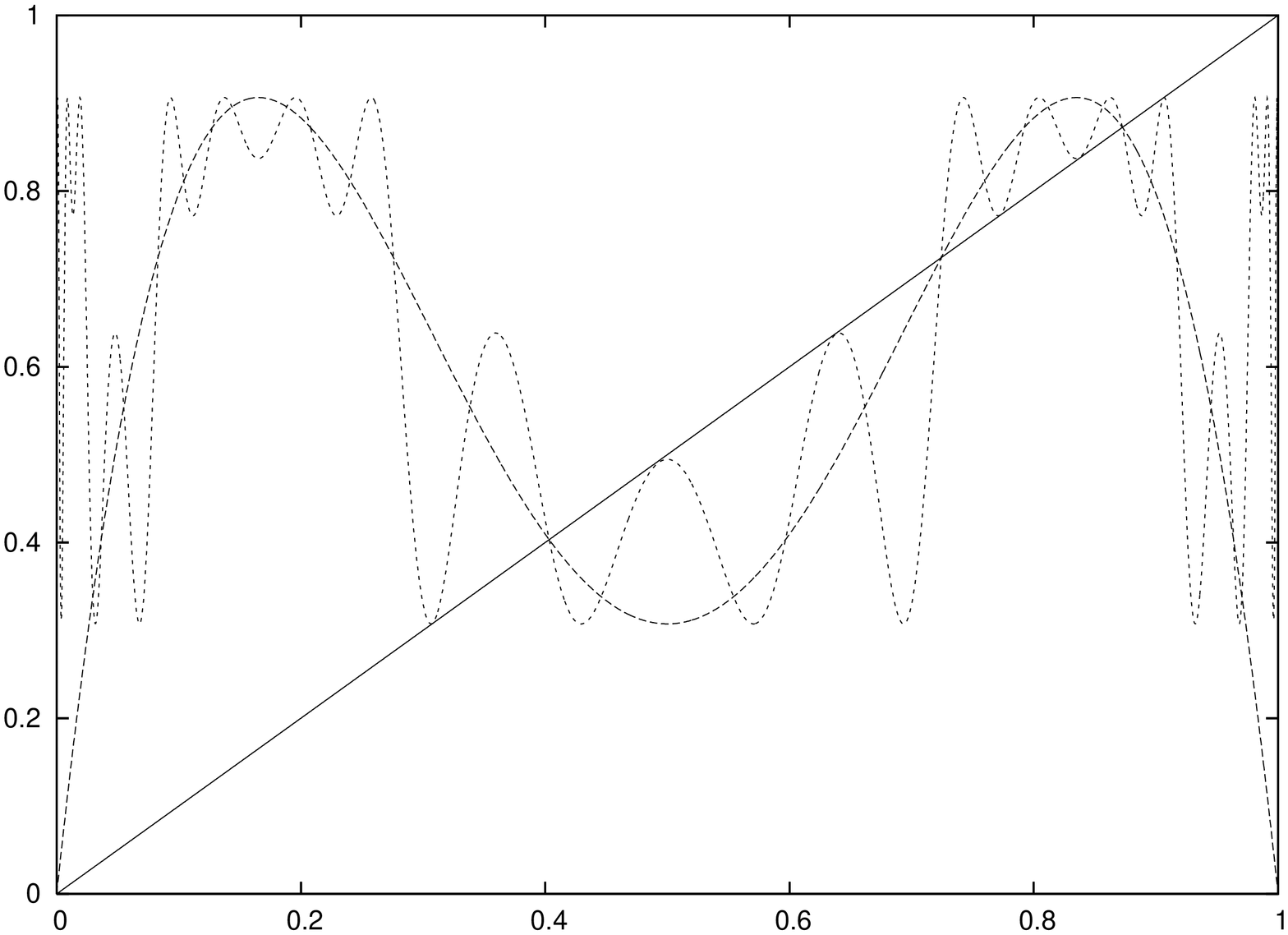}\end{center}
\end{figure}

As shown in Fig. \ref{cap:fig1} we can see a saddle-node bifurcation
of $f^{3}$. This same figure appears twice in Fig. \ref{cap:fig2}.
They are the two first elements of the saddle-node bifurcation cascade
of $f^{3\cdot2^{n}}$. The bigger $n$ is the more times Fig. \ref{cap:fig1}
is replicated in the graph of $f^{3\cdot2^{n}}$ along the line $y=x$. 

As shown in the appendix, Fig. \ref{cap:fig1} appears twice more
every time we move on one stage in a saddle-node bifurcation cascade,
half of the figures are contracted by $\frac{1}{\alpha}$ and the
other half by $\frac{1}{\alpha^{2}}$. The outcome is that in a saddle-node
bifurcation cascade the points, that are tangent to the line $y=x$
, duplicate at the same time as the region they are placed in contracts.

We would hope to find any kind of Cantor set and, what is worse, one
Cantor set for each period-$q\cdot2^{n}$, $q\neq2^{m}$ saddle-node
bifurcation cascade, because all saddle-node bifurcation cascades
approach the Myrberg-Feigenbaum point $r_{\infty}$ as $n\rightarrow\infty$.
Nonetheless the solution is extraordinary simple at the limiting value
$r_{\infty}$.

If we consider the cascade $q,q\cdot2,...,q\cdot2^{n},...$, $q\neq2^{m}$
it will turn out that the graph of $f^{q}$will be reproduced in the
neighborhood of the critical points of $f^{\,2^{n}}$, which correspond
to the points of the restricted-supercycle given by $\left\{ \frac{1}{2},f(\frac{1}{2}),f^{2}(\frac{1}{2})....,f^{2^{n}-1}(\frac{1}{2})\right\} $(see
appendix). Half of these neighborhoods are contracted by $\frac{1}{\alpha}$
and the other half by $\frac{1}{\alpha^{2}}$, each time we move on
one stage in saddle-node bifurcation cascade, that is, the saddle-node
points duplicate. Therefore, in the limit $n\rightarrow\infty$each
neighborhood has collapsed to a point of $\left\{ \frac{1}{2},f(\frac{1}{2}),f^{2}(\frac{1}{2})....,f^{2^{n}-1}(\frac{1}{2})\right\} _{n\rightarrow\infty}$.
The points of $\left\{ \frac{1}{2},f(\frac{1}{2}),f^{2}(\frac{1}{2})....,f^{2^{n}-1}(\frac{1}{2})\right\} _{n\rightarrow\infty}$
coincide with the period doubling orbit as $n\rightarrow\infty$.
The outcome is that the period-$2^{n}$ ( $n\rightarrow\infty$) chaotic
orbit coincides with period-$q\cdot2^{n}$, $q\neq2^{m}$, $n\rightarrow\infty$
orbit ---in the sense of limit---, that is, the limits cannot tell
from each other. We have the same limit orbit, both the one which
comes from period doubling cascade at $r<r_{\infty}$, and the one
which comes from saddle-node bifurcation cascades at $r>r_{\infty}$. 

The fact that the period-$q\cdot2^{n}$saddle-node orbit tends to
the period-$2^{n}$ as $n\rightarrow\infty$hides another fact: the
collapse of period-$q\cdot2^{n}$window to a point at Myrberg-Feigenbaum
point $r_{\infty}$. This is so because the period-$q\cdot2^{n}$
window starts with the birth of the period-$q\cdot2^{n}$saddle-node
orbit and the period-$q\cdot2^{n+1}$ window starts with the birth
of the period-$q\cdot2^{n+1}$saddle-node. As the birth of both saddle-node
orbits tends to $r_{\infty}$, as $n\rightarrow\infty$, then the
window length tends to zero. This result is captured in the expression
(see \cite{SanMartin05})\[
\frac{L_{n}}{L_{n+1}}=\delta\]
 which shows that the length of two successive windows of a saddle-node
bifurcation cascade are contracted by a factor $\delta$, being $\delta$
Feigenbaum constant, that is, the windows length tends to zero.

If we consider the period-$q\cdot2^{n}$window it will have a Myrberg-Feigenbaum
point $r_{\infty,q\cdot2^{n}}$, at which its corresponding Feigenbaum
cascade will finish. As $n\rightarrow\infty$, the window length tends
to zero and it brings the following consequences and interpretation,
relative to period-$q\cdot2^{n}$window\@.

i) The whole period doubling process also collapses to a point, and
the same happens to the rest of the saddle-node bifurcation cascades
present in the period-$q\cdot2^{n}$ window, because the process occurs
in a window whose length tends (collapses) to zero.

ii) The Myrberg-Feigenbaum point $r_{\infty}$ of canonical window
and the Myrberg-Feigenbaum point $r_{\infty,q\cdot2^{n}}$of the period-$q\cdot2^{n}$
window are the closer to each other the bigger $n$ is. The distance
tends to zero as $n\rightarrow\infty$. The same happens with every
one of saddle-node bifurcation cascades located in the period-$q\cdot2^{n}$
window. Accordingly the accumulating point of every saddle-node bifurcation
cascade (a new Myrberg-Feigenbaum) tends to $r_{\infty,q\cdot2^{n}}$,
and therefore it tends to $r_{\infty}$. The process is applied again
to the new windows which are born from a saddle-node bifurcation cascade
and so forth.

This explains why a Myrberg-Feigenbaum point is an attractor of other
Myrberg-Feigenbaum points, which are attractor of other Myrberg-Feigenbaum
points and so forth (see \cite{SanMartin05})

The former convergence process has been expounded for a fixed saddle-node
bifurcation cascade with $q\neq2^{m}$, but it is valid for any value
of $q$. Therefore there are infinite sequences which mimic the former
process, one for each value of $q\neq2^{m}$.

The approaching to the Myrberg-Feigenbaum point, and the multiplicity
of convergent sequences, explain completely the mechanism of attractor
of attractor introduced in \cite{SanMartin05}

\section{CONCLUSIONS}

The presence of saddle-node bifurcation cascade in the logistic equation
assures the genesis of a intermittency cascade. Each saddle-node bifurcation
of the bifurcation cascade is associated with an intermittency. As
the location of the saddle-node bifurcation is known it brings that
so is the genesis of the intermittency cascade. The knowledge, a priori,
of the length of the laminar regime in the type-I intermittency, and
of the scaling the peaks and valleys of the successive iterated of
logistic map, allows us to establish the length of the laminar regime
in the intermittency cascade.

The intermittency cascade is a phenomenon that takes place in all
windows of the logistic map, and not only in windows associated to
first-occurrence orbits.

It is proved that the windows collapse to the Myrberg-Feigenbaum points,
this mechanism being responsible for the fact that Myrberg-Feigenbaum
points are attractors of attractors.

\section*{Acknowledgments}

The author wishes to thank Daniel Rodríguez-Pérez for helpful discussions
and help in the preparation of the manuscript.

\appendix

\section*{APPENDIX}

Let's find the scaling law of high-order cycles in the saddle-node
bifurcation cascade. To do so we will follow the Feigenbaum work \cite{Feigenbaum80}
, which is carried out close to pitchfork bifurcation.

The scaling law is not determined by the location of the elements
on the x-axis, but by their order as iterates of $x=\frac{1}{2}$
( or of $x=0$ after a coordinate translation that moves $x=\frac{1}{2}$
to $x=0$). This is the point which necessarily belongs to any supercycle.
Feigenbaum denotes the distance from the m-th element of a $2^{n}$-supercycle
to its nearest neighbor by\[
d_{n}(m)=x_{m}-f_{R_{n}}^{2^{n-1}}(x_{m})\]
where $R_{n}$is the control parameter value at which supercycle occurs.

To generalize the latter definition to the period-$q\cdot2^{n}$ saddle-node
orbit, which is in the saddle-node bifurcation cascade, we are not
going to take into account all its points, but only a few very particular
ones.

As a period-$q\cdot2^{n}$ saddle-node orbit undergoes a period doubling
process there will be a control parameter value, prior to the duplication
, at which the orbit will be a period-$q\cdot2^{n}$ supercycle. Let
$R_{n,q}$ be such parameter, and let be\begin{equation}
\left\{ \frac{1}{2},f(\frac{1}{2}),....,f^{q\cdot2^{n}-1}(\frac{1}{2})\right\} \label{s1}\end{equation}
the supercycle in question.

We extract from supercycle the sequence\[
\left\{ x_{m,q}\right\} _{m=1}^{m=2^{n}}=\left\{ x_{1,q}=\frac{1}{2},x_{2,q}=f(\frac{1}{2}),x_{3,q}=f`^{2}(\frac{1}{2})....,x_{2^{n},q}=f^{2^{n}-1}(\frac{1}{2})\right\} \]
which we will name {}``restricted supercycle''. The restricted supercycle
consist of $2^{n}$ points, for every one of which the function $f^{2^{n}}$has
a critical point close to the line $y=x.$ The restricted supercycle
is similar to the supercycle with which Feigenbaum works in the period
doubling cascade, but it is different because there are $q$ points
of the supercycle (\ref{s1}) around every critical point close to
line $y=x$. Furthermore,the neighborhood of every critical point
is visited $q$ times if the order given by supercycle (\ref{s1})
is followed. In other words, the graph of $f^{q}$is in the neighborhood
of every point of a restricted supercycle. Accordingly, the scaling
law of restricted supercycle gives the scaling law of the supercycle
(\ref{s1})

\begin{figure}[ppp]

\caption{\label{cap:fig4}Graph of $f^{3\cdot2^{2}}$(thin dashed line), at
Saddle-Node bifurcation, reproduces the graph of $f^{3}$ (see Fig.
\ref{cap:fig1}) around $f^{2^{2}}$ (thick dashed line) maxima and
minimum.}

\begin{center}\includegraphics[%
  width=0.70\textwidth,
  keepaspectratio,
  angle=-90]{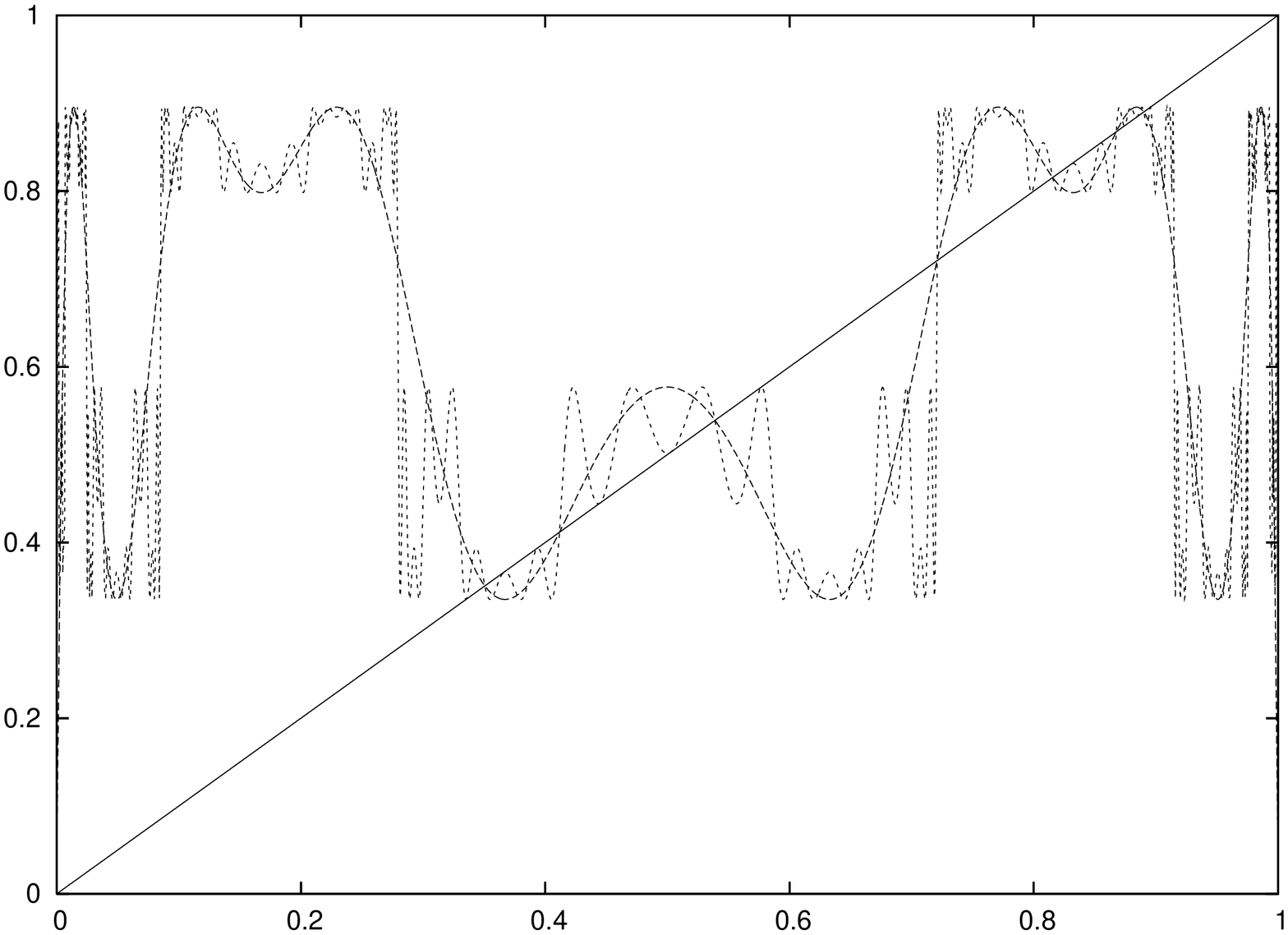}\end{center}
\end{figure}

What we are doing is to classify the $q\cdot2^{n}$ points of the
saddle-node orbit in $2^{n}$sets, each one having $q$ points. The
$2^{n}$ sets correspond to the $2^{n}$ critical points of $f^{2^{n}}$closest
to line $y=x$. The neighborhood of each one of these critical points
captures $q$ saddle-node points, in other words, the graph of $f^{q}$
is captured in every one of such neighborhoods; for instance, in $f^{3\cdot2^{2}}$we
notice that the graph of $f^{3}$ is captured in the neighborhoods
of the $2^{2}$ critical points of $f^{2^{2}}$ (Fig. \ref{cap:fig4}).

Let's denote the distance from the m-th element of a $2^{n}$-restricted-supercycle
to its nearest neighbor by

\[
d_{n,q}(m)=x_{m,q}-f_{R_{n,q}}^{2^{n-1}}(x_{m,q})\]

Let's define the scaling (see Eq. (56) of \cite{Feigenbaum80}) by 

\[
\sigma_{n,q}(m)=\frac{d_{n+1,q}(m)}{d_{n,q}(m)}\]

Bearing in mind that $x_{m}=f_{R_{n,q}}^{m}(0)$, if we set $m=2^{n-i}\,\,\,\,1\ll i\ll n$
then $\sigma_{n,q}(2^{n-i})$ can be approximated by (see 57 of \cite{Feigenbaum80})

\[
\sigma_{n,q}(2^{n-i})\sim\frac{g_{i+1,q}(0)-g_{i+1,q}\left[(-\alpha)^{-i}g_{1,q}(0)\right]}{g_{i,q}(0)-g_{i,q}\left[(-\alpha)^{-i+1}g_{1,q}(0)\right]}\]
where $g_{i,q}$ are the functions defined in \cite{SanMartin05}.

The new variable

\[
t_{n}(m)=\frac{m}{2^{n}}\]
 or \[
t_{n}(2^{n-i})=2^{-i}\]
rescales the axis of iterates in such a way that all $2^{n+1}$ iterates
are within a unit interval.

Defining 

$\sigma_{,q}(t_{n}(m))\sim\sigma_{n,q}(m)$ (as $n\rightarrow\infty$)
\\
it turns out that (see Eq. (60) of \cite{Feigenbaum80})

\[
\sigma_{,q}(-2^{-i-1})=\frac{g_{i+1,q}(0)-g_{i+1,q}\left[(-\alpha)^{-i}g_{1,q}(0)\right]}{g_{i,q}(0)-g_{i,q}\left[(-\alpha)^{-i+1}g_{1,q}(0)\right]}\]

In the limit $i\rightarrow\infty$ it yields

\[
\sigma_{,q}(-2^{-i-1})_{i\rightarrow\infty}=\frac{g(0)-g\left[(-\alpha)^{-i}g_{1,q}(0)\right]}{g(0)-g\left[(-\alpha)^{-i+1}g_{1,q}(0)\right]}=\frac{1}{\alpha^{2}}\]
where 

\[
g_{i+1,q}(x)\rightarrow_{i\rightarrow\infty}g(x)\]
has been used (see \cite{SanMartin05}), and besides $g(x)$ has a
quadratic maximum, so 

\[
g\left[(-\alpha)^{-i}g_{1,q}(0)\right]\simeq g(0)+\frac{1}{2}g^{''}(0)\cdot(-\alpha)^{-2i}g_{1,q}^{2}(0)\]

Let's notice that the scaling law does not depend on $q$, so we can
drop this label and simply write

\[
\sigma(-2^{-i-1})_{i\rightarrow\infty}=\frac{g(0)-g\left[(-\alpha)^{-i}g_{1,q}(0)\right]}{g(0)-g\left[(-\alpha)^{-i+1}g_{1,q}(0)\right]}=\frac{1}{\alpha^{2}}\]

The independence of $q$ is critical to expound that the whole set
of Saddle-Node bifurcations (for any $q$) scales as the set of pitcfork
bifurcation described by Feigenbaum. Once $\sigma(-2^{-i-1})$ has
been calculated, Feigenbaum \cite{Feigenbaum80} gives numbers in
binary expression and demonstrates that $\left|\sigma\right|$ behaves
as $\sim\frac{1}{\alpha}$ half the time and as $\sim\frac{1}{\alpha^{2}}$
the other half. 

\begin{thebibliography}{10}

\bibitem{Feigenbaum78} M. J. Feigenbaum, Quantitative Universal for a Class of Nonlinear
Trans\-formations. Journal of Statistical Physics, Vol. 19, 25 (1978)
\bibitem{Feigenbaum79} M.~J.~Feigenbaum, The Universal Metric Properties of Nonlinear Transformations. Journal of Statistical Physics, 21 669-706 (1979)
\bibitem{Ruelle71} D. Ruelle and F. Takens, {}``On the Nature of Turbulence''. Commun.
Math. Phys. 20, (1971) 167-92
\bibitem{Newhouse78} S. E. Newhouse, D. Ruelle and R. Takens, {}``Occurrence of Strange
Axiom A Attractors near Quasi-Periodic Flows on $T_{m}$ ($m=3$ or
more)''. Commun. Math. Phys. 64, 35 (1978)
\bibitem{Manneville79} P. Manneville and Y. Pomeau, {}`` Intermittency and the Lorenz Model''.
Phys. Lett. 75A (1979) 1-2
\bibitem{Pomeau80} Y. Pomeau and P. Manneville, {}``Intermittent transition to turbulence
in dissipative dynamical systems''. Commun. Math. Phys. 74 (1980)
189-197
\bibitem{Anil97} C. V. Anil Kumar, T. R. Ramamohan, New Class I intermittency in the
dynamics of periodically forced spheroids in simple shear flow. Phys.
Lett. A 227 (1997) 72-78
\bibitem{Bauer92} M. Bauer, S. Habip, D. R. He, W. Martienssen, New type of intermittency
in discontinuous maps. Phys. Rev. Lett. 68 (1992) 1625-1628
\bibitem{SanMartin99} J. San-Martín and J. C. Antoranz, {}``Type-I and Type-II Intermittencies
with Two Channels of Reinjection''. Chaos, Solitons \& Fractals Vol.
10 N. 9 1539-1544(1999)
\bibitem{SanMartin05} J. San-Martín, Universal Scaling in Saddle-Node Bifurcation Cascades (I). [nlin.CD/0501035]
\bibitem{Gukenheimer87} J. Gukenheimer, {}``Renormalization of one dimension mappings''
Contemp. Math 58, pt. III (1987) 143-160
\bibitem{Feigenbaum80} M. J. Feigenbaum, Universal Behavior in Nonlinear Systems. Los Alamos
Sciences 1 4-27 (1980)

\end{thebibliography}
\end{document}